# Modeling the Stellar Spectral Energy Distributions of Star-Forming Galaxies

Claus Leitherer

*Space Telescope Science Institute, 3700 San Martin Drive, Baltimore, MD 21218*
*leitherer@stsci.edu*

**Abstract.** I will review recent progress in the modeling of the stellar spectral energy distributions of star-forming galaxies. I will cover the full relevant wavelength range from the near-infrared to the extreme ultraviolet, with an emphasis on the ultraviolet long- and shortward of the Lyman break where most of the stellar luminosity is emitted. Uncertainties in stellar atmosphere and evolution models will be critically examined, and the impact on the total panchromatic luminosity will be highlighted.

## INTRODUCTION

M82 is the archetypal dwarf starburst galaxy. A plethora of observational data over many decades of the electromagnetic spectrum has led to a comprehensive understanding of the starburst phenomenon in this galaxy. It was only 40 years ago when the first steps were made towards recognizing the true nature of M82. The pioneering study of [3] related the velocity field of the extended gas to the central explosion and to the presence of unseen O and B stars in the vicinity of the nucleus. The population of OB stars has eluded direct detection since then, but panchromatic observations of the spectral energy distribution (SED) of M82 have allowed us to draw a fairly accurate picture of the underlying processes. In Fig. 1 I have reproduced the overall SED from the Balmer limit to the sub-mm regime. Three distinct components stand out: thermal emission from warm dust peaking around 50 μm, discrete emission features around 10 μm caused by vibrations and out-of-plane bending of polycyclic aromatic hydrocarbon (PAH) molecules, and direct stellar light in the optical and near-infrared (IR). The stellar component is the subject of this review. It is the powering source of the emitted energy at all wavelengths. The source of the luminosity longward of ~10 μm in Fig.1 is absorbed and re-radiated stellar light. In the absence of dust absorption, the stellar part of the SED would continue rising shortward of 1 μm, with the difference between the attenuated and unattenuated SED being roughly equal to the mid- and far-IR emission in Fig.1.

In the following sections, I will review stellar spectra, atmospheres, and evolution models, which are the fundamental building blocks. Combined with synthesis models, they allow construction of the SED. Predicted quantities such as colors, lines indices, and spectra can be tested against observations and cases of success and failure can be identified.

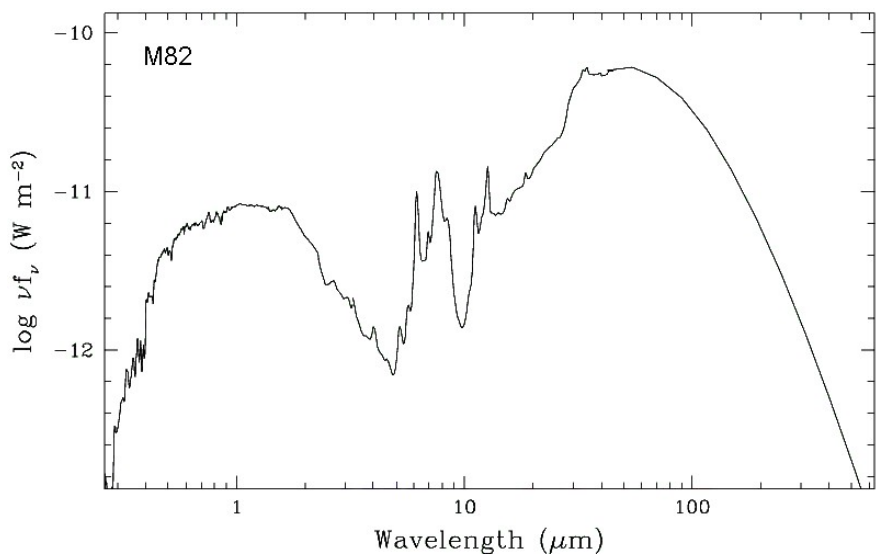

**FIGURE 1.** Panchromatic SED of M82 from the Balmer limit to the sub-mm regime (from [19]).

## MODEL COMPONENTS

The ingredients determining the observed *stellar* part of a galaxy SED are the spectra of individual stars, prescriptions for superposing them, plus effects that cannot be accounted for from first principles. This section addresses the latest developments related to spectral libraries and the construction of realistic synthetic populations from evolution models and a stellar initial mass function (IMF). I will also comment on rarely discussed, but potentially very important, morphological effects.

### Stellar Spectra

Galaxies are made of stars. Stellar spectra are the basic building blocks of a galaxy SED. A stellar spectral library can be either empirical or theoretical ([34]). In the past, models were often limited by the amount of available computer time, lack of atomic data, or the complexity of the physics. On the other hand, dust reddening and the restrictions imposed by the available spectral windows often necessitate a theoretical approach. Recently, there has been a shift in preference away from empirical to theoretical libraries. The main reasons are the reliability of the latest generation of model atmospheres and the need to cover the full parameter space, which would otherwise be observationally inaccessible. Nearby stars have the Galactic chemical evolutionary history imprinted, and their spectra are therefore sometimes badly suited for comparisons with those of, e.g., elliptical ([47]) or Lyman-break galaxies ([40]).

STELIB is a new *empirical* stellar library compiled by [28]. STELIB consists of a homogeneous library of several hundred stellar spectra in the visible range (3200 to 9500 Å) at intermediate spectral resolution (~3 Å). This library includes stars of various spectral types and luminosity classes, spanning a broad range in metallicity. The spectral resolution, wavelength and spectral type coverage of this library represent a substantial improvement over previous libraries used in population synthesis models. It is implemented in the latest version of GISSEL ([2]).

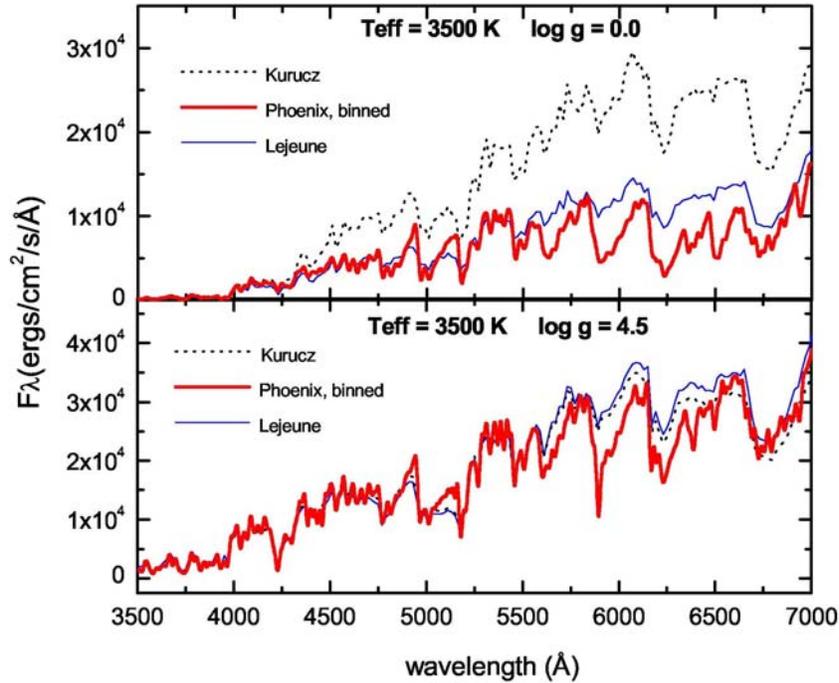

**FIGURE 2.** Comparison of the fluxes predicted by the models of [22], [33], and [16]. The latter are based on Phoenix atmospheres and predict the overall shape in a self-consistent manner. Upper panel: supergiants; lower panel: dwarfs ([37]).

Current *theoretical* efforts are reflected in the synthetic library of [37] who computed ~1600 high-resolution stellar spectra, with a sampling of 0.3 Å and covering the wavelength range from 3000 to 7000 Å. The library was computed with the latest improvements in stellar atmospheres, incorporating non-LTE, line-blanketed TLUSTY models ([25]) for hot, massive stars and spherical, line-blanketed Phoenix models accounting for tri-atomic molecules ([16]) for cool stars. The grid covers the full Hertzsprung-Russell diagram (HRD) for a wide range of chemical abundances. The replacement of the traditional Kurucz ([22]) atmospheres at both the hot and cool end is significant. For instance, important diagnostics such as the helium lines are strongly affected by non-LTE effects in O stars. LTE models predict too weak equivalent widths. An example of the corresponding improvement at low temperature ($T_{eff}$) is shown in Fig. 2. Phoenix models self-consistently account for the overall SED because of their extensive molecular line list. These molecular transitions are missing in the widely used Kurucz models. [33] provided an empirical correction to the Kurucz models. This correction becomes unnecessary with the new fully blanketed models.

The ultraviolet (UV) part of the spectrum is emitted by hot stars. The strong winds in hot stars require modeling with spherically extended, expanding non-LTE atmospheres. Above the Lyman limit, these models can account for the observed strong stellar-wind lines related to highly ionized metals such as, e.g., Si IV λ1400 or

C IV λ1550 ([31]). The wavelength range below 912 Å is of greater relevance to the topic of this conference: it is the source of the radiative heating of the interstellar gas. The dramatic effects of non-LTE and sphericity effects are demonstrated in the newly released model atmosphere set of [46]. In Fig. 3, these models ("WM-basic") are contrasted with classical, static LTE atmospheres ("Kurucz") and with CoStar models, which are similar to WM-basic models, except for the exclusion of line-blanketing. WM-basic is considered to have the best physical ingredients and is the atmosphere of choice for photoionization modeling. All models agree rather well in the wavelength range where the $H^0$ ionizing radiation is emitted ($\lambda > 504$ Å); the number of photons predicted to be emitted in the hydrogen Lyman continuum is a robust quantity which has changed by less than 0.2 dex since the classical work of [42]. The behavior of the ionized helium continuum below 228 Å is in sharp contrast to that of hydrogen. Now, wind effects drastically alter the spectrum, depending on blanketing and wind density. For solar chemical composition (top panel), high wind densities lead to $He^+$ recombination and a marked drop of the flux. Lower composition leads to weaker winds, and $He^+$ remains ionized. In this case, a vast flux excess over Kurucz models is predicted, yet blanketing still lowers the output below that of the CoStar models.

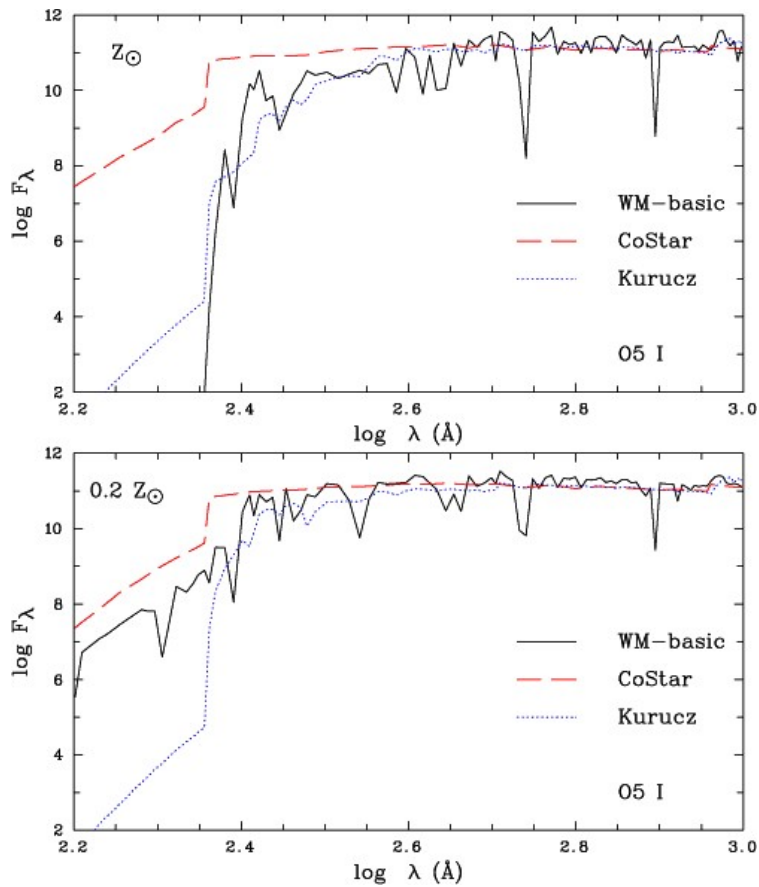

**FIGURE 3.** Comparison of the emergent fluxes for an early O supergiant from a WM-basic (solid), a CoStar (dashed), and a Kurucz model (dotted) at solar metallicity and 0.2 Z ([46]).

The overall state of stellar spectral libraries, both empirical and theoretical, is rather satisfactory. Current efforts are pushing for an extension of the available parameter space towards low and non-standard chemical composition, an area of interest to cosmologist. The available spectral templates are still scarce but rapidly growing.

## Stellar Evolution Models

Construction of composite SEDs from individual stellar spectra requires prescriptions for the appropriate weighting by the stellar temperatures, luminosities, and lifetimes. Stellar evolution models provide such prescriptions. Excellent reviews on the subject can be found in [6] and [36]. As the most luminous stars dominate the composite SED, it is primarily massive-star evolution that is relevant. Stellar evolution in the upper part of the HRD mainly depends on the chemical composition at birth, the stellar mass and its decrease with time due to mass loss, and mixing processes induced by rotation, convection, and overshooting.

The main focus of massive-star evolution models has shifted from core overshooting in the 1980s, over mass loss in the 1990s, to rotation in the new millennium. These processes have one theme in common: they determine (among others) the relative importance of the convective vs. the radiative energy transport. The higher efficiency of convection has many effects on stellar evolution, most notably on the ratio of lifetimes spent in the core hydrogen-burning stage and in the shell hydrogen-burning stage. The impacts of such effects on the composite SED have been discussed by [53]. These processes are effective only for stars massive enough to develop convective cores, therefore they are visible only in young and/or massive populations, which then display bluer spectra.

Evolution models with rotation are just beginning to become available ([18]). The new models demonstrate the significance of rotation: The additional helium brought near the H–burning shell by rotational mixing and the larger He–core both lead to a less efficient H–burning shell and a smaller associated convective zone. Therefore, the stellar radius of rotating stars will inflate during the He–burning phase. A pilot computation for a 20 M star is reproduced in Fig. 4. Rotation can become the principal parameter for post-main sequence evolution if stellar rotation velocities on the main sequence are above ~200 km s$^{-1}$, values not unrealistically high.

The largest uncertainty is the unknown behavior of stellar rotation at very low metallicity. Circumstantial evidence for higher stellar rotation velocities in the Magellanic Clouds was presented by [35]. Should this trend continue to extremely low metallicities, rotationally induced convection processes could be very important, e.g., in Population III stars. If so, metals could be transported to the surfaces of the first generations of stars, inducing stronger winds, higher opacity, and a significantly altered extreme UV radiation field. The consequences for the chances of detection of these stars would be quite significant.

The uncertainties in stellar evolution modeling primarily affect highly evolved stars, either in the red (red supergiants [RSG]) or blue (Wolf-Rayet [W-R] stars) part of the HRD. On the other hand, the vast majority of the stellar population contributing to the composite SED is fairly well understood and reproduced with state-of-the-art stellar evolution models.

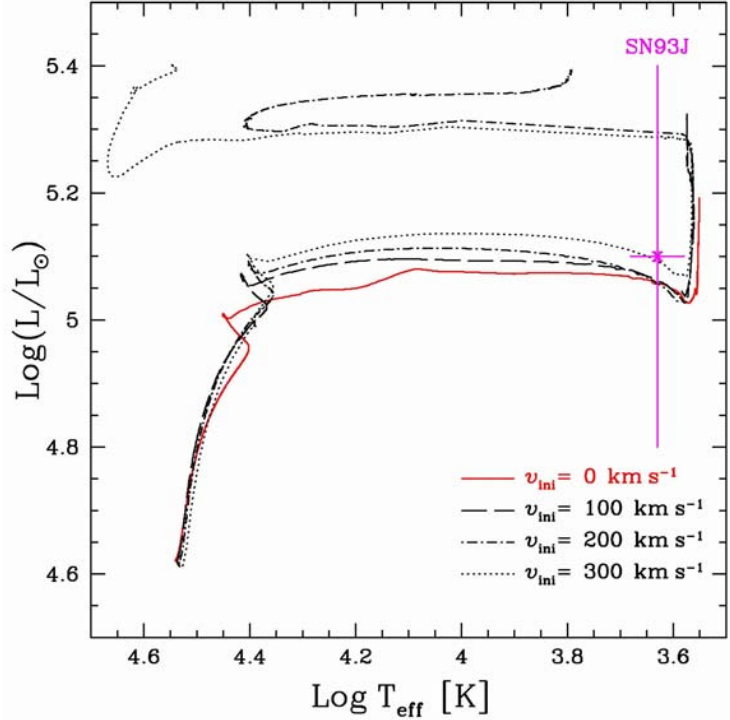

**FIGURE 4.** Evolutionary models for a 20 M☉ star: solid, dashed, dotted-dashed, and dotted lines correspond to rotation velocities of 0, 100, 200, and 300 km s$^{-1}$, respectively ([18]).

## Stellar IMF

The upper part of the HRD is degenerate in terms of the relation between stellar parameters and mass, even for a single stellar population. The situation is very different from, e.g., a globular cluster isochrone where most of the luminosity comes from a small mass interval close to the turn-off mass. In a typical massive-star population of single age, stars with vastly different zero-age-main-sequence masses can have similar $T_{eff}$ and $L$ and contribute to the integrated light. The SED is therefore dependent on an assumption on the IMF because of the existence of a steep mass-to-luminosity relation in the upper HRD.

The royal way of deriving an IMF for young populations is via clusters. This is feasible only in few cases so that IMF data are often uncertain and restricted to a limited mass range ([38]). Since massive stars are rare and have much larger luminosities than low-mass stars, it is difficult to observe the full stellar mass spectrum in one and the same star cluster. The low-mass star content of the closest massive-star formation region in Orion has been studied down to ~0.1 M☉. This region, however, contains only a handful of stars with masses above 10 M☉. In contrast, the R136 cluster in the LMC is rich enough for meaningful statistics above 10 M☉, but it is difficult to push the low-mass star detection limit below 1 M☉. The IMF in nearby star-forming galaxies has been derived using various indirect observational techniques using the integrated light. In these cases, the accessible stellar masses are between 100 M☉, where stochastic effects due to small-number statistics set in, and 1 M☉, where velocity dispersion measurements are used ([26]).

Almost all studies suggest a rather uniform IMF in star-forming galaxies. Considering the very different observational techniques used and the prevailing theoretical uncertainties, this result suggests that any existing variation and its dependence on the environment are smaller than the current measuring uncertainties. The classical Salpeter IMF and its modern variant defined by Kroupa appear to provide the best match to the data ([21]).

## Dust Morphology

A major complication for interpreting SEDs of gas- and dust-rich galaxies is the potentially large uncertainty introduced by dust obscuration. Star-forming galaxies are associated with large molecular gas densities. This immediately suggests that dust must be plentiful since molecular hydrogen forms by adsorption on dust grains. Moreover, star formation comes together with metal production. Therefore star formation itself will rapidly enrich the ISM in heavy elements and dust. This occurs over rather short time scales. The time scales for the formation of dust are only a few hundred Myr ([7]) so that obscuration effects can set in fairly early in the history of a galaxy.

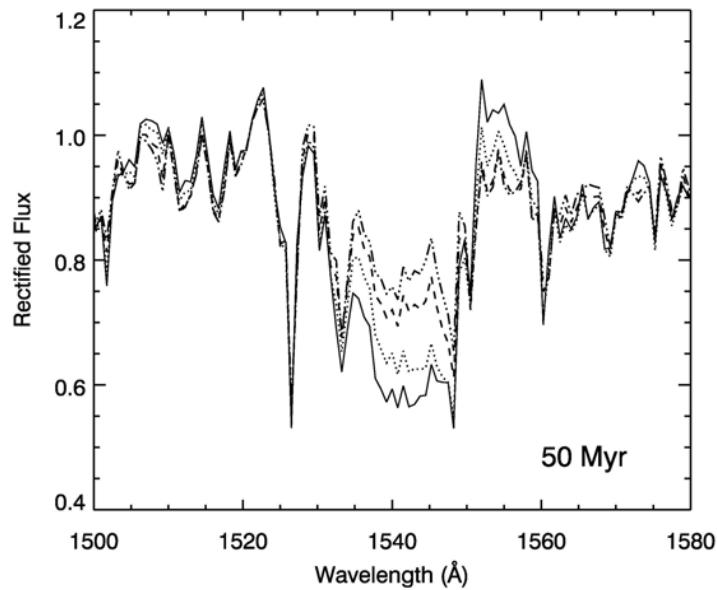

**FIGURE 5.** C IV $\lambda 1550$ for a model of a 50 Myr old stellar population forming stars constantly. The different line types denote different dust dispersal time scales. Solid line: no time dependence; dotted line: 5 Myr; dashed line: 15 Myr; dash-dotted line: 50 Myr ([29]).

Dust efficiently scatters and absorbs UV radiation. In order to quantify the interaction between dust and photons, assumptions on the amount of dust must be made, as well as on its geometry and chemical composition. Three processes play a role. (i) Dust *obscures* matter, i.e., dust conceals stars and gas from view by covering them wholly or in part. (ii) Dust *attenuates* light, i.e., it diminishes the amount of light seen by an observer. This leaves open the possibility of either absorption or scattering. (iii) Dust *absorbs* photons, i.e., it transforms the photons energy. The significance of these processes becomes obvious by recalling that in the Local Group the UV extinction law varies from galaxy to galaxy and within galaxies. The extinction is due to a combination of absorption and scattering of photons. The geometry is quite different in more distant galaxies where individual stars cannot be isolated. In such cases much of the light may be scattered into the line of sight, and the properties of the dust become paramount for the interpretation of the observed photon distribution ([4]).

These dust effects are of course always accounted for before making comparisons with observations. What is often not appreciated, is the subtle interplay of the evolution of the stellar SED and the spatial morphology of the dust. In Fig. 5 I have plotted models for the important C IV $\lambda$1550 line, which assume time-dependent dust attenuation. This line is formed by a population of OB stars, with the emission and absorption coming from the most massive O stars, and the continuum from less massive O, and some B stars. The reason for the "dilution" of the line profile over time is a gradual decrease of the dust reddening with time. The astrophysical justification is that we are seeing the less massive stars through a hole where the energetic O-star winds have blown out the natal cocoon veiling the most massive stars. Very young clusters are known to remain embedded in their natal material until energetic stellar winds from evolving massive stars blow out the surrounding gas and dust ([50]). If younger populations are preferentially more attenuated, the *equivalent widths* of C IV $\lambda$1550 and other lines become skewed towards smaller values. If the ISM is inhomogeneous, a fundamental property of a spectral line associated with a single star changes: *its equivalent width becomes reddening dependent*.

This ISM structure invalidates the frequently made assumption of isotropy and homogeneity. If different stellar phases are associated with different dust columns, the composite SED becomes dependent on the dust morphology and its evolution, even for the assumption of a simple foreground screen model.

## PREDICTIONS

Combining a suitable set of spectra with evolution models allows the synthesis of spectrophotometric quantities for an assumed population mixture, a concept that dates back to [49]. Such models describe the spectral and chemical evolution of stellar systems in an attempt to derive the properties of a stellar population in both nearby and distant galaxies, whose individual stars are either resolved or unresolved, respectively. A recent summary of the different modeling approaches and results can be found, e.g., in [32] and [2].

The time evolution of the global SED of a single stellar population is shown in Fig. 6 ([51]), whose purpose is to highlight both the dramatic change of the SED with time and how different stellar evolution models affect the predictions. The comparison is performed for the most widely used evolution models by the Geneva [41], original Padova [11], and Padova models extended beyond the asymptotic giant branch (AGB) phase [51]. Young population with ages of 1, 3, 10, 50, and 100 Myr are shown in the left panel. SEDs produced with the Padova tracks are more luminous by up to 0.1 dex at most wavelengths. These differences are the result of higher intrinsic stellar luminosities in the Padova models for post-main-sequence evolutionary phases. There is a striking difference between the SEDs in the UV around 3 Myr. Hot W-R stars when coupled with extended model atmospheres produce the excess emission in the Padova models.

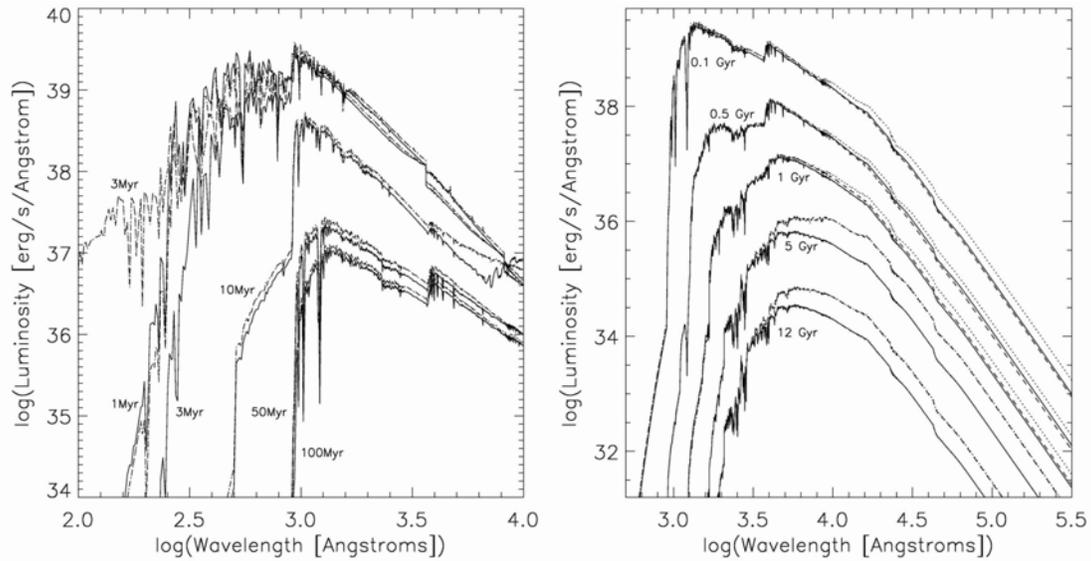

**FIGURE 6.** Spectral energy distributions for young (left) and old (right) populations with solar composition. Solid: Geneva; dashed: original Padova; dotted: Padova with AGB stars added. The far-UV continuum produced by the Geneva and Padova models at 3 Myr differs dramatically and has been labeled twice for clarity. Note that the Padova and the Padova+AGB models coincide prior to the occurrence of the first AGB stars, producing a "dashed-dotted" graph ([51]).

Older ages of 0.1, 0.5, 1, 5, and 12 Gyr are reproduced in the right panel of Fig. 6. The contribution of AGB stars to the red part of the spectrum is clearly visible. For models with ages >1 Gyr (e.g., 5 Gyr), there is a drastic difference between the Padova and Geneva tracks. The difference results from a combination of two effects. First, the Geneva tracks do not include low-mass stars below 0.8 M. Since the chosen IMF extends down to 0.1 M, the absence of these stars creates a luminosity deficit, which sets in at an age of a few Gyr, and becomes progressively stronger. Second, the implementation of the Geneva tracks in the Starburst99 models used in Fig. 6 does not include the add-on set published by [5]. [5] computed the horizontal-branch evolution after the He-flash in low-mass stars. Horizontal-branch stars around ~1.5 M contribute significantly to the near-IR luminosity, beginning at an age of ~1 Gyr. They are the chief reason for the difference between Geneva and Padova after that epoch.

The main trends in Fig. 6 are (i) a gradual shift of the peak wavelength from the blue to the red due to the disappearance of massive stars and the appearance of red giants; (ii) more pronounced line-blanketing and associated flux depression in the blue than in the red; (iii) stronger variations over time in the blue than in the IR. Next, I will turn to *colors* in order to quantify these effects and to provide direct comparisons with observations.

In Fig. 7, I illustrate the influence of different stellar evolution models on the predicted photometric evolution for solar metallicity. Shown are models computed using the Padova 1994 ([8], [9]), the Geneva ([41]), and the Padova 2000 ([11]) tracks. *(B – V)* and $M/L_V$ are quite similar with the three sets of tracks, whereas *(V – K)* shows rather significant differences. This reflects the good agreement of the model predictions for stars relatively close to the main-sequence, which influence the optical colors, and the larger uncertainty of the giant and RSG colors in the near-IR. The largest differences arise at early ages (< 10 Myr) because of the larger number of evolved W-R stars in the Padova models than in the Geneva models. Another notable difference exists between ages of 10 and 20 Myr when RSGs dominate in the near-IR. AGB stars become important between $10^8$ and $10^9$ yr, and the differences in the color evolution reflect uncertainties of the AGB evolution. Furthermore, since the minimum mass for quiet helium ignition is lower in the Geneva than in the Padova 1994 models (1.9 M versus 2.2 M ), the helium flash occurs at slightly later ages in the Geneva models (see the bump around 1 Gyr in Fig. 7). At late ages, the *(V – K)* color is significantly bluer in the Padova 2000 model than in the Padova 1994 model. The reason is the 50 – 200 K higher $T_{eff}$ of the red giant branch in the 2000 versus the 1994 tracks.

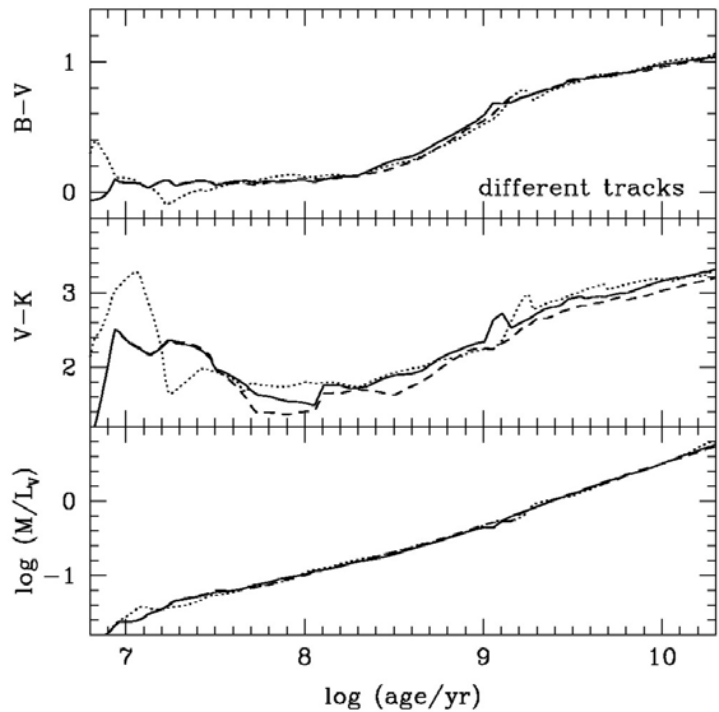

**FIGURE 7.** Evolution of the *(B – V)* and *(V – K)* colors and the stellar mass-to-light ratio $M/L_V$ of simple stellar populations of solar metallicity computed using the Geneva (dotted line), Padova 1994 (solid line), and Padova 2000 (dashed line) stellar evolution prescriptions ([2]).

Colors are subject to multiple degeneracies caused, among others, by metallicity, age, or reddening. If permitted by the observations, one would always want to compare data and models at the highest possible resolution. In Fig. 8, I am showing the spectral evolution of a single stellar population with the same parameters and using the same technique as in Fig. 6. The spectra were calculated from *model atmospheres*. The difference between the SEDs in the two figures is the 100 times higher spectral resolution in Fig 8 (0.2 Å). This resolution permits diagnostics beyond those based on colors. For instance, Fig. 8 highlights the evolution of the Balmer series limit until its disappearance around ~1 Gyr, followed by the formation of the 4000 Å break. Such diagnostics can be successfully traded to break the previously mentioned degeneracies.

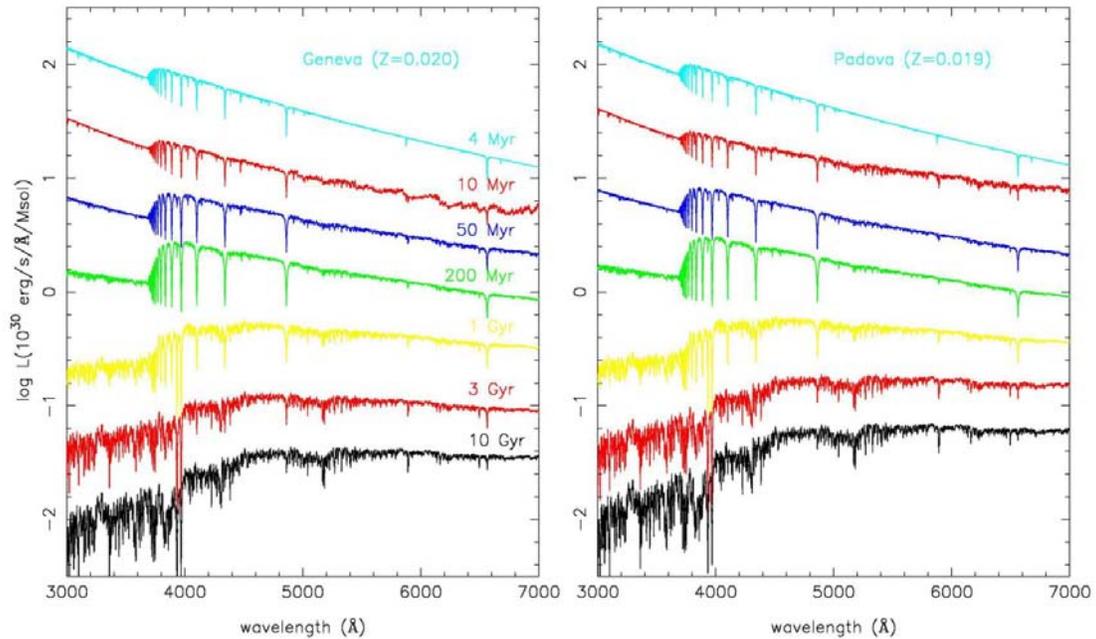

**FIGURE 8.** Theoretical spectral evolution of a single stellar population with solar composition predicted by the Geneva (left) and the Padova 2000 (right) models. Ages from top to bottom are 4 Myr, 10 Myr, 50 Myr, 200 Myr, 1 Gyr, 3 Gyr, and 10 Gyr ([12]).

An alternative approach is to incorporate *empirical spectral libraries* into a synthesis code and perform suitable interpolations in parameter space for complete HRD coverage. This is the method used by PEGASE-HR, a new stellar population synthesis program generating high-resolution spectra at $R = 10,000$ in the optical range between 4000 and 6800 Å ([27]). The program links the spectrophotometric model of galaxy evolution PEGASE.2 ([10]) to an updated version of the ELODIE library of stellar spectra. An example is reproduced in Fig. 9. [27] used these new high-resolution spectra to calculate sets of line indices for age- and metallicity determinations. Some of them are based on the classical Lick indices ([52]). The high spectral resolution opens up a new dimension in parameter space: $R$ is large enough to address kinematic effects. Kinematic studies require the model resolution to be significantly higher than the resolution of the data as imposed by the galaxy velocity dispersion. With $R = 10,000$, even low-luminosity dwarf galaxies or even very luminous stellar clusters become accessible to modeling.

## OBSERVATIONAL TESTS

How reliable are the models? In this section I will compare the models to data, starting from a morphologically simple giant H II region, and then proceeding to large star-forming galaxies. The purpose is to identify evolutionary phases that work as well as those still needing improvement.

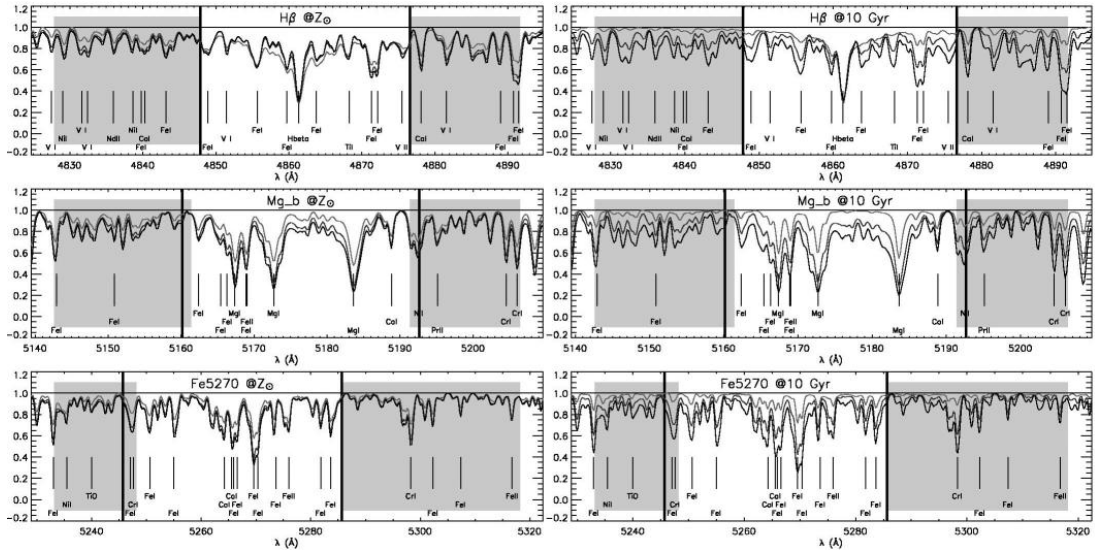

**FIGURE 9.** Evolution of the principal stellar lines as a function of age for ages of 1, 4, 13 Gyr (from light gray to black, respectively) at $Z_\odot$ (left) and metallicity [Fe/H] = –1.7, –0.4, 0.4 (from gray to dark, respectively) at 10 Gyr (right). Gray areas show the blue and red pseudo-continua of the Lick indices and solid vertical lines delimit the central passbands ([27]).

NGC 604 in M33 is the second most luminous H II region in the Local Group of galaxies, after 30 Dor in the LMC. Its proximity (0.85 Mpc) permits detailed studies of both individual stars and its integrated properties. Thus NGC 604 is an excellent test case for theoretically predicted SEDs. [13] performed a self-consistent analysis of the UV-to-optical spectrum of NGC 604 (Fig. 10, left panel). By combining space-UV and ground-based optical spectroscopy and imagery, they identified the individual energetic components of the stellar population and measured their balance. The far-UV below 912 Å is not accessible to direct observations but the emitted photons of hot stars are absorbed in the outer atmospheres and used to accelerate powerful stellar winds. Typically about 30% of the available radiative momentum is converted into kinetic momentum ([23]). Observational signatures of the wind are strong UV lines of, e.g., O VI $\lambda$1035 or C IV $\lambda$1550, which are broadened and blueshifted by up to 2000 km s$^{-1}$. The wind features are strong enough to be dominant in any H II region spectrum. [30] discussed how the strength and in particular the blueshift of the UV lines can be used to infer the age and mass spectrum of the underlying stellar population. The basic concept is the tight relation between stellar-wind properties (i.e., the line shape and velocity) and stellar luminosity, which is an immediate consequence of a radiatively driven wind. This method has become the standard tool to study stellar populations in the UV and was used by [13] to infer the nature of the stars powering NGC 604.

The modeling of the UV in turn provides a complete prediction for the ionizing spectra of the stellar population in NGC 604. This energy distribution can then be used as input for photoionization models of the nebular emission lines observed at longer wavelengths. Photoionization models are an alternative, independent constraint on the massive-star content, therefore providing clues on systematic uncertainties inherent in the synthesis modeling. In the case of NGC 604 (and in almost all other cases studied so far), both approaches lead to the same answer.

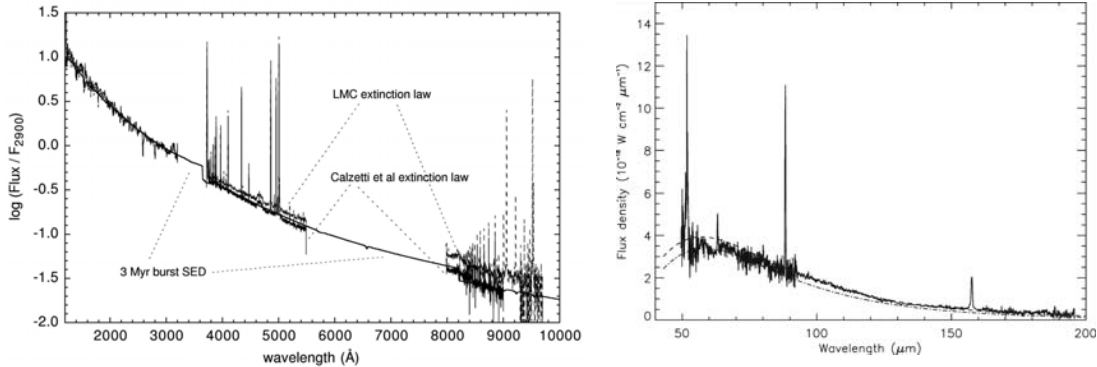

**FIGURE 10.** Left: Model for a 3 Myr old instantaneous burst, normalized at 2900 Å, compared with ultraviolet plus optical spectra of NGC 604. The data were dereddened with $E(B-V) = 0.1$ using the LMC extinction law, and with $E(B-V) = 0.2$ using the extinction law of [4]. Adapted from [13]. Right: ISO LWS grating spectrum of NGC 604. Two gray-body continuum models are shown: $T = 42$ K with $\alpha = 1.0$ (dashed line) and $T = 38$ K with $\alpha = 1.5$ (dot-dashed). Adapted from [17].

The emission-line modeling can be pushed one step further by including the near- and mid-IR part of the spectrum where prominent fine-structure lines are located. These lines have several key advantages: (i) Different ionization stages of the same element are at similar wavelengths; (ii) reddening corrections are almost non-existent; (iii) the exponential term of the volume emission coefficient vanishes for IR fine-structure lines because of their low excitation energy. Therefore, IR lines have weak temperature dependence, and it is possible to study nebular parameters free of concerns over temperature fluctuations. [17] analyzed ISO LWS spectra of NGC 604 and used the fine-structure lines observed in the far-IR to model the H II and photodissociation regions. The derived densities and ionization fractions agree with those found in the optical. Furthermore, the mid- and far-IR dust emission is consistent with the attenuation inferred from the optical. The ISO spectrum of NGC 604 is bounded by two models, either with a temperature $T = 38$ K, $\alpha = 1.5$, and a visual extinction $A_V = 0.9$ mag, or with a temperature $T = 42$ K, $\alpha = 1$, and $A_V = 0.0$ mag. ($\alpha$ is the power-law exponent of the dust opacity.) The corresponding continuum models are compared to the ISO data in the right panel of Fig. 10. For comparison, the UV and optical data suggest that the reddening affecting the stellar cluster is $E(B-V) = 0.1$, if the LMC attenuation law is adopted, or $E(B-V) = 0.2$ if a Calzetti-type law is used. The SED is well matched with either of these attenuation laws. This attenuation value is also similar to the average absorption of $A_V = 0.5$ derived for the gas from radio observations ([13]).

While theoretical SEDs compare rather favorably with simple systems, like star clusters and H II regions, large systems, such as galaxy nuclei remain a challenge. [24] combined UV to near-IR data to constrain the young, the intermediate-age, and the old stellar populations in the central region of the starburst galaxy NGC 7714. The size scale is a factor of 10 larger than in NGC 604's case. The young burst responsible for the UV light represents only a small part of an extended episode of enhanced star formation, which began a few times $10^8$ yr ago. NGC 7714 owes its brightness in the UV to a few low extinction lines of sight toward young stars. The different extinction values obtained when different population indicators are used result naturally from the coexistence of populations with various ages and obscurations. The near-IR continuum image looks smoothest, as a consequence of lower sensitivity to extinction and of a larger contribution of old stars. The global SED is the result of the averaging over many lines of sight with very diverse properties in terms of obscuration and stellar ages. Ultimately, strong degeneracies preclude *detailed* modeling of the stellar light from the near-IR to the space-UV in the majority of galaxies.

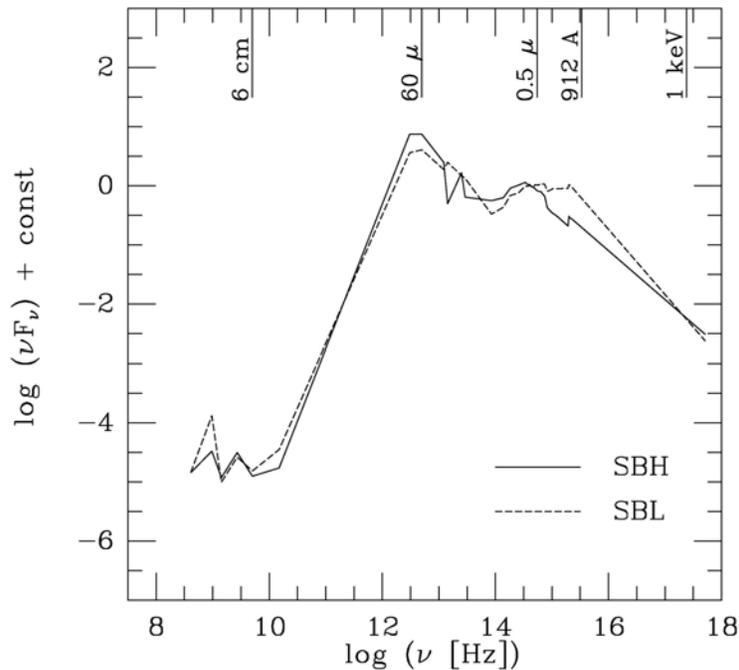

**FIGURE 11.** Average SEDs from the radio to X-rays for high- (SBH) and low-reddening (SBL) star-forming galaxies (from [44]). The spectra are normalized such that $\log(\nu F_\nu) = 0$ at 5500 Å.

Ignoring such "microphysics", I will now assume a global viewpoint and provide an empirical description of the average, energy distribution of the local star-forming galaxies from the radio to X-rays. The emphasis will be on global properties of the spectrum and its relation to the stellar and interstellar components of the galaxy.

[44] used a predominantly UV selected sample of 26 star-forming galaxies to construct average, global energy distributions. The average SEDs between 6 cm and 10 keV are shown in Fig. 11. The sample is subdivided in high- ($E(B–V) > 0.4$) and low-reddening ($E(B–V) < 0.4$) galaxies. Both samples have similar overall spectral characteristics, except in the far-IR and the non-ionizing UV, which are anti-correlated: objects with lower reddening have higher UV and lower far-IR flux, whereas the opposite holds for objects with higher reddening. The dispersion of the average curves around the peak wavelengths is quite small, indicating a rather uniform

behavior. The interpretation is obvious: star-forming galaxies with larger reddening have higher dust content and are more effective in absorbing UV photons and re-emitting the processed radiation in the far-IR. The flux difference between the two data sets at 1500 Å is about a factor of 5 to 10 and corresponds to the escape fraction of non-ionizing UV radiation.

The photons observed in the spectra of galaxies with active star formation originate either in stars or in the ISM. Starlight dominates from the near-IR to the far-UV (2.2 μm to 912 Å). All other parts of the energy distribution in Fig. 11 are due to interstellar dust (60 μm) and gas (6 cm and X-rays). The entire spectrum is of course powered by the stellar energy input alone, and the radio and X-ray radiation have their origin in the non-radiative stellar luminosity by winds and supernovae (SN). Fig. 11 can be used for a rough estimate of the ratio between the radio- and X-ray luminosity over the bolometric luminosity, assuming the peak at about 60 μm is indicative of $L$. The average star-forming galaxy has $L_{X-ray}/L \approx 10^{-3}$ and $L_{radio}/L \approx 10^{-5}$.

SED models predict a very tight relation between the UV and the bolometric luminosity for star-forming populations in equilibrium. *In this case, the non-ionizing UV luminosity (912 – 3650 Å) accounts for about 75% of the total luminosity, with a negligible dispersion. 20% are emitted in the Lyman continuum, with the remaining 5% contributed by the Paschen continuum.* These relations and their small dispersions explain of the complexity of modeling the optical-to-near-IR SED of a star-forming galaxy: only a small fraction of the luminosity is sampled, and wrong assumptions on the dust attenuation and additional, unaccounted stellar components can become the dominant factors. In contrast, the small dispersion of the mean ionizing and non-ionizing luminosities, together with the high ISM opacities at these wavelengths, are the prerequisite for the validity of star-formation tracers that are not affected by second order effects: nebular recombination radiation and thermal dust emission in the far-IR.

[20] investigated the Hα and far-IR star-formation rate diagnostics for 81 galaxies in the Nearby Field Galaxy Survey. There is a strong correlation between the ratio of the two star-formation indicators (Fig. 12). This correlation suggests that the gas and dust must be closely coupled in all far-IR-bright galaxies. The most plausible explanation for this effect is that the majority of the dust and gas heating in galaxies occurs very close to the actively star-forming regions. Modeling by [43] for normal star-forming galaxies strengthens this explanation. Radiation from H II regions is predicted to be the dominant energy source for the dust emitting at 60 μm. Even at 100 μm, H II regions contribute about one third of the radiation source. The other two third is produced by the diffuse UV and optical radiation, which could also contain some contribution from the young stellar population.

Even if the young stellar population dominates the far-IR emission, one would expect the ratio of gas and dust masses to vary from one H II region to another and from one galaxy to another. Similarly, the dust geometry and composition must vary. If these variations occur, they do not appear to have an effect on the relationship between the global far-IR and Hα luminosities. It is similarly intriguing that the correlation holds for both early and late galaxy types. The IR emission from early-type spiral galaxies may contain a significant contribution from the general stellar radiation field. This contribution should increase the observed IR luminosity. If this effect were offset by lower dust opacity, a remarkable conspiracy would be required to hold for all galaxy types. It seems more plausible that the far-IR emission results from the same young stellar population that produces the Hα emission in all galaxy types. This hypothesis is also supported by other physical correlations linking the far-IR emission to the young stellar population, including the radio-far-IR correlation.

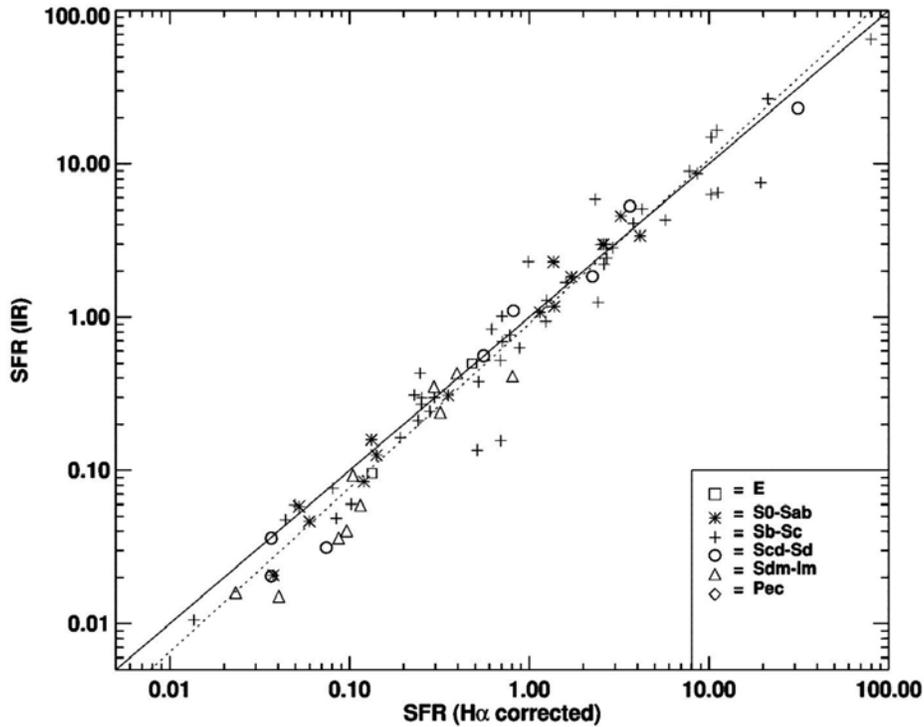

**FIGURE 12.** Comparison of the star-formation rates derived from the IR and Hα. The strong correlation spans four orders of magnitude. The solid line is y = x and shows where the data would lie if both indicators agreed. Hα has been corrected for extinction. The dashed line is the best fit to all the data. The legend indicates the Hubble type (from [20]).

The tight correlation of Fig. 12 also requires the IMF to be fairly uniform. Stars producing the bulk of the ionizing radiation have characteristic masses of ~50 M , whereas the non-ionizing continuum is emitted by stars with masses of ~20 M . The fact that the two star-formation indicators agree so well excludes any strong variation of the upper IMF. Furthermore, Fig. 12 is an empirical confirmation of the validity of several modeling assumptions, including predictions for the ionizing radiation field of massive stars.

I will conclude this review with an open issue. Overall, stellar modeling is rather reliable, and the more so for stars close to the main sequence. One remaining nagging issue concerns RSGs. I will elaborate on this trouble spot using blue compact dwarfs (BCDs) as an example. BCDs are thought to be strongly starbursting galaxies powered by ionizing stars and with an underlying population of red stars whose precise age is still under discussion (e.g., [1]; [48]). There is, however, convincing evidence for a significant number of RSGs in those objects with well-established CMDs ([45]).

[51] collected near-IR photometry of a sample of BCDs from the literature and compared them to synthesis models. The large-aperture photometry is sensitive to the young starburst, the surrounding field consisting of potential earlier starburst episodes, and the older underlying population. Therefore any RSGs present in these galaxies will affect or even dominate the photometry. The observations collected from the literature are plotted and compared to models in Fig. 13. Both reddening and line emission are negligible at IR wavelengths. The synthetic colors predicted for a single stellar population (SSP) and for continuous star formation are in the left and right panels of Fig. 13, respectively. Fig. 13 (left) suggests reasonable agreement between the bulk of the data and the computed colors of a SSP with solar chemical composition.

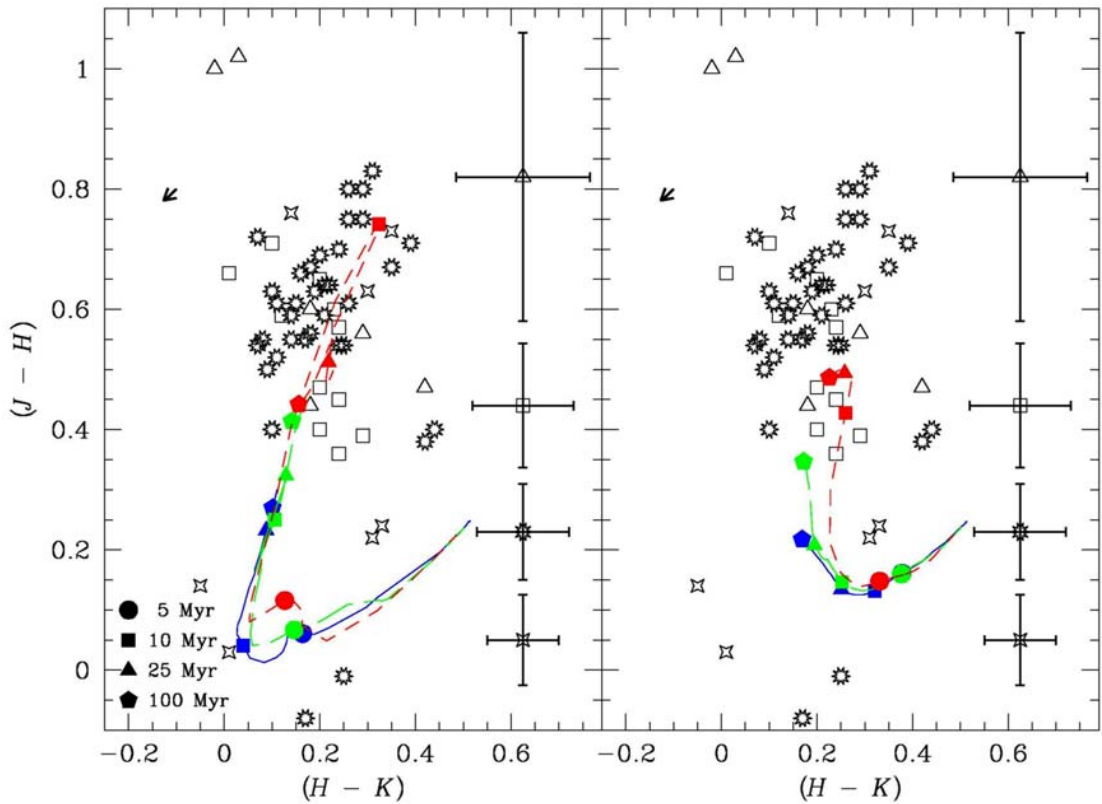

**FIGURE 13.** Color-color diagrams of four samples of BCDs compared with starburst models of $Z$ = 0.001 (solid line), 0.004 (long-dashed), and 0.02 (short-dashed line) based on the Geneva tracks. Filled symbols indicate the ages of the models. Open symbols show the data points collected from the literature. The error bars are the dispersion value in each sample. The solid vector indicates the reddening correction for $A_V$ = 0.25. Left: instantaneous star formation; right: continuous star formation ([51]).

The synthetic models for SSPs were terminated at an age of 100 Myr. Higher ages are unrealistic, as BCDs are defined via their emission lines and blue colors. 100 Myr old starbursts would not be classified as BCDs anymore. The details of the star formation during the first tens of Myr, however, are a subject of debate. Dwarf galaxies are known for their complex star-formation histories, with periods of quiescence and intermittent bursts of star formation ([15]). Depending on the burst frequency, the effective star formation may mimic a steady-state situation. This is addressed in the right panel of Fig. 13. As expected, the imprint of the RSGs is more diluted, and the colors are more degenerate than for a SSP. If the star-formation history in BCDs were constant, the comparison between the data and the models would force one to postulate ages far in excess of 100 Myr. While this would not necessarily be in conflict with observational selections (ionizing stars are continuously replenished), the associated gas consumption would become entirely unreasonable. BCDs do not constantly form stars over 1 Gyr. Therefore the appropriate star-formation scenario is between the extremes plotted in Fig. 13, but most likely much closer to the SSP case in the left panel.

Does this suggest consistency between the observed and synthetic colors? The only track in Fig. 13 (left) that matches the data points is the one at solar chemical composition. The other tracks at lower abundance are significantly bluer and fail to reproduce the observed colors regardless of the assumed age and reddening correction. Only the solar models produce RSGs in large enough numbers and with sufficiently low Teff to reach the colors covered by the data points. Yet, the approximate average oxygen abundance of the sample is 20% solar. Therefore the assumption of solar composition is invalid, and the applicable models are those with $Z = 0.004$. Once forced to compare the data to the $Z = 0.004$ tracks, one arrives at the inescapable conclusion that the predicted colors of evolutionary models for metal-poor populations with a significant RSG component are incorrect. This conclusion is unchanged for either the Padova or Geneva tracks. Our results echo those of [39], who demonstrated that both the Geneva and Padova evolution models fail to predict the location of RSGs and the Large and Small Magellanic Clouds. The evidence of failure at solar chemical composition is much weaker, if present at all.

## THE FUTURE

The wavelength region of interest for studying important physical processes tends to fall outside the domain accessible with ground-based telescopes. This applies to both the mid-IR and the space-UV. Past generations of space instruments could not afford the resolution required by the relevant astrophysical spatial scales. The IR in particular suffered from this shortcoming, as the IRAS mission had very limited spatial resolution, and fully panchromatic SED studies were restricted to the spatially integrated light.

With ISO and Spitzer, we have entered a new era. Panchromatic imagery of M81 over six decades in wavelength space is shown in Fig.14 ([14]). The proximity of M81 has made it a favorite target for many investigations of galaxy properties from the X-ray to radio. Many well-resolved images have been taken in the past, with the notable exception of the IR. Previously, well-resolved far-IR images of galaxies have only been possible for Local Group galaxies such as the Magellanic Clouds. With the successful launch of Spitzer, it is now possible to map many large galaxies in the far-IR with good spatial resolution, good sensitivity, and in a reasonable amount of time. M81 is one of the key galaxies in the Spitzer Infrared Nearby Galaxies Survey ([14]).

Two of the questions that can be probed with these new observations are the variation of star-formation indicators and the IR-radio correlation across the disk of M81. MIPS, Spitzer's mid-IR imager, permits comparisons with UV and Hα images to probe the behaviors of the IR, Hα, and UV star-formation indicators across M81. Such comparisons have been made for *global* galaxy fluxes ([20]), but rarely has it been possible to resolve all three indicators in a single galaxy. By studying their resolved behavior, we will greatly improve the accuracy of these star formation indicators for both resolved and global galaxy measurements.

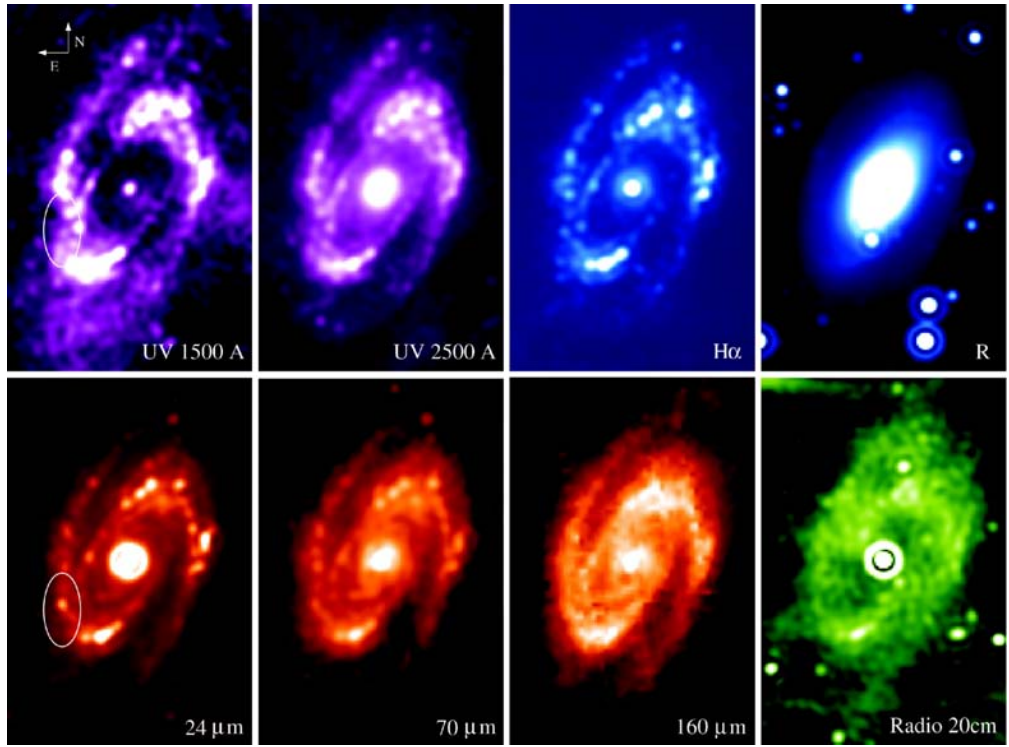

**FIGURE 14.** Panchromatic view of M81 using Spitzer, UIT, and ground-based images at wavelengths of 1500 Å, 2500 Å, Hα, R, 24 μm, 70 μm, 160 μm, and 20 cm. The field of view of the images is 15″ × 23″ ([14]).